\newenvironment{namelist}[1]{%
\begin{list}{}
    {
      
      \settowidth{\labelwidth}{#1}
      \setlength{\leftmargin}{1.1\labelwidth}
    }
  }{%
\end{list}}

\documentstyle[epsf,preprint,twoside,adbis95%
,amscd%
,times ]{acmconf} 
\author{V.E.Wolfengagen\thanks{also:
Kashirskoe Avenue, 31, Cybernetics Department,
Moscow Engineering Physical Institute,
Moscow, 115409, Russia
}
              \vspace{1.52mm} \\
{Vorotnikovsky per., 7, bld. 4} \\
{Institute for Contemporary Education ``JurInfoR-MSU''} \\
{Moscow, 103006, Russia} \\
{\tt vew@jmsuice.msk.ru}
}
\addtocounter{footnote}{1}
\title{Object-oriented solutions
\thanks{\em This research is supported by the
Russian Foundation for Basic Research
(project 93-012-943)}%
}
\begin{document}
%%%%%%%%%%%%%%%%%%%%%% Bibliographystyle  %%%%%%%%%%%%%%%%%%%%%%%%
\bibliographystyle{alpha}
%%%%%%%%%%%%%%%%%%%%%%%%%%%%%%%%%%%%

\maketitle

%%%%%%%%%%%%%%%%%%%%% Abstract %%%%%%%%%%%%%%%%%%%%%%%%%

\begin{abstract}
In this paper are briefly outlined the motivations,
mathematical ideas in use, pre-formalization and
assumptions, object-as-functor construction, `soft' types
and concept constructions, case study for concepts based
on variable domains, extracting a computational background,
and examples of evaluations.
\end{abstract}

%%%%%%%%%%%%%%%%%%%%%%%%%%%%%%%%%%%%%%%%%%%%%%%%%%%%%%%

%%%%%%%%%%%%%%%%%%%%%%%%%%%%%%%%%%%%%%%%%%%%%%%%%%%%%
%%%%%%%%%%% The body of the contribution %%%%%%%%%%%%
%%%%%%%%%%%%%%%%%%%%%%%%%%%%%%%%%%%%%%%%%%%%%%%%%%%%%

\section{Introduction}
% \addcontentsline{toc}{section}{Introduction}

An early incite in a {\em theory of computations} was to
incorporate
objects for a variety of purposes. They were assumed to represent
the
existent - {\em actual, possible or virtual} objects in a
problem domain. The nature of existence was also under the
concentrated study. The recent years have generated a lot of object
assumptions and discussions.
Nevertheless, the initial notion of an object
became overloaded by the mismeaning and not significant features.
Every new research in the area added the excessive troubles to
understand the clear sense and meaning of the object paradigm.

An attempt to rearrange the useful ideas will be done here.
The main attention is paid to establishing the parallelism
between a theory of computations and the object-oriented
notions.

\subsection{Motivation for object evaluator}

Object can be represented by embedding in a host computational
environment. An embedded object is accessed by the laws
of the host system. A pre-embedded object is observed as
the decomposition into substitutional part and access function
part which are generated during the object evaluation.
They assist to easy extract of the result.

Subsumption is an usual theory-of-computation technique.
Counterparts of the entire method~-- logic, functor category,
and applicative computations,~-- are attached to generate
an intermediate computational framework. This intermediate
representation is indirectly based on the categorical combinatory
logic. The needed optimizations may be obtained equationally.

The resulting model seems to be a kind of {\em object evaluator}.
The object evaluator feature is to incorporate the schematic
elements which are subdivided into {\em individuals} and
{\em individual concepts}. Both of the entities are based on
the notion of the {\em variable domain}. This is a schematic
construction and is equipped with both the cloning and
transactional means to capture {\em dynamics}.

All the parts of object evaluator share the same functor model
with the parameterized types and assignments. The logical part
has been supplied with both the atomic and non-atomic formulae
with the variables ranging over the variable domains. The
categorical part assists the evaluation to enable the
extraction of a substitutional part. The applicative part
is capable of separating the computation paths for function
and its argument.

In this paper are briefly outlined the motivations,
mathematical ideas in use, pre-formalization and
assumptions, object-as-functor construction, `soft' types
and concept constructions, case study for concepts based
on variable domains, extracting a computational background,
and examples of evaluations.

\subsection{Evolution of the ideas}
% \addcontentsline{toc}{subsection}{Evolution of the ideas}

A technical intuition for an object is approximately as follows:
object is proclaimed to be an entity (by default) with the
strictly attached attributes~-- `internal state' and
`behavior'~\cite{EhGoSe:93}. Some of the objects are called the
`dynamic objects', that communicate with each other (note that
{\em communication} is presupposed of great importance). Next step
is to classify objects by their type, to collect objects into
classes, to superimpose the various inheritances, to compose for
generating complex objects. Note that computational intuition
tends to establish object as a mathematical process.

\subsubsection{Logic to incorporate objects}

An approach to apply logic to a phenomena of object
seems to be clear and natural (e.g., \cite{HeCo:89}, \cite{Jacob:92}
\cite{Ga:93}). Nevertheless, adoption
of more or less traditional approach of logic is distant
of the essence of the initial task to be solved.
When the researcher was to pick
this kind of science it would combine some significant elements. \\
1. The conditions of reasoning that transcend not only logic,
but both the mathematics and the theory of computation(s). \\
2. The traditions of observation and insight that led into
the foundations of these sciences.

The semantics of traditional logics is often used but this
argument is not sound.  The most important for the notation
is to be usable by the computational tool that
applies it to the {\em environment} to produce the {\em results}.

To share the concern for the rigorous theory it is not
necessary to adopt all the amount of any particular
formalism. The more prominent approach seems to be based
more on the constraints that can be superimposed by
the problem. If the existing formalism turns off to match these
conditions then that means a perspective to find out for
meaningful thing.

If to confine the search for the theory of objects to areas were
formalism has already succeed in the answer may to be missed. A
necessary theory is likely to be found where the logic meets the
incompleteness, troubles of intensions etc. As a rule, the
traditional logical machinery seems to be well applied to
pre-formalized reality and is not suitable equipped with the means
for more dynamic occasions. When we go back to the generic
principal ideas we have more possibilities to expand the
predefined tools to deal with the problem as it is arose and used
by the computational devices.

\subsubsection{Manifesting a category theory}
% \addcontentsline{toc}{subsection}{Manifesting a category theory}

An early trouble was the suitability of a theory for
the working researcher. The same is for a category theory.

The theoretician position (see \cite{Lawv:75}, \cite{Go:90},
\cite{EhGoSe:91})
seems to have embraced
category theory as the pre-eminent
universal science, to adopt its more or less traditional approach
with a possible missing the significant initial features.

The term `arrow thinking' as it is used in a category theory
refers to the standardized notion~-- within this theory,~-- that
prescribes a mapping of the terms and expressions of the initial
system into a world of abstract functions. But it is only one
element of categorical philosophy. In most significant
applications of category theory such a thinking does not map
symbolic expressions into real objects with the substantial
properties, and such models become only imaginable.

For some systems of logic model is described by the theory,
e.g. in the form of cartesian closed category (c.c.c.).
The need is to manipulate the elements.
A domain $T$ is said to have an element if there is a map
$h:I\to T$ (here: $I$ is a domain of assignments).
If $f$ is a constant function $f:T\to S$, then
$f\circ h:I\to S$. Thus, maps in c.c.c. can behave as
functions on elements.

\subsubsection{Applicative computational systems}

A lot of theories (not necessary logic or category theory)
have the ultimate goal to develop a notion or construction
which suits for the interpretation of abstract objects.
For instance, {\em $\lambda-$calculus} and {\em combinatory logic}
contain in their foundation a concept of object to suit
the computational needs (\cite{Sco:80}).

Moreover, an isomorphism between intuitionist logics and typed
$\lambda-$calculi was established. An original Curry's theory of
functions generated formula-as-type notion under which a proof of
a statement $\exists x B$ is a pair $<a,b>$ consisting of an
object $a$ and a proof $B[a]$. In practice, a {\em type} is
regarded as an abstract object whereas a {\em formula} is the name
of a type.

All of this is in harmony with a principal feature of the
applicative computations, namely: (1)~the symbols of function
and its argument are treated separately as the distinct objects;
(2)~the first object is applied to the second under an
{\em application} metaoperator. The advantages of this approach
are not yet entirely observed.

\subsubsection{Intermediate theoretical framework}

All of the theories above seem to have an universality.
The method is to add the restrictions to enrich the pure
theory by the needed sensitivity.

For instance, the connection between $\lambda-$calculus and c.c.c.
(\cite{CoCuMa:85})
has generated the variants of categorical combinatory logic.
A basic concept for the approach was given by the set of
abstract objects, namely, {\em categorical combinators}.
This kind of objects is within both a category and computational
system. They share the clear advantages of the
distinct subsystems.

\subsubsection{Introducing abstract objects}

A `phenomena' of object was discussed many times with
a lot of attitudes. Some selected and superimposed
questions seem to be as follows:

\begin{itemize}
\item[] how new individuals come into existence and go away
as situation changed;
\item[] how concepts get their semantics in realistic conditions
e.g., with a tremendous set of possible worlds;
\item[] traditional (logical) machineries are usable to prove
the existence of an individual (under some properties) but give no
equipment to name that, possibly generated, individual and refer
to it by name in further consideration;
\item[] first-order logic provides a tool to support
the necessary truths and their consequences. It provides
no machinery to deal with the relationships with
the creation or destruction of individuals;
\item[] what is the machinery to characterize the way
individuals change over time;
\item[] what is an ability to reify the notion of the state
(in different contexts);
\item[] how to talk about both the direct and side effects
of the actions;
\item[] $\dots$
\end{itemize}

All of this place the state-of-things in the proper
perspective.
All of this clearly indicate that
the long term hoped-for unified logic,
categorical framework, or computational system
is not yet reached.
The variety of logics, theories, models, and approaches
tends to more growth.

\section{Restricting the topic: pre-formalization}

Some efforts to encircle the task will be needed.
Both direct and indirect solutions are substantially
based on putting the ideas in a certain order.
Subsumption is a common technique shared by distinct
`dimensions'~-- logical, categorical, and computational
(\cite{Wo:93}).

\subsection{The starting assumptions}

Most of the approaches start with the notion of a {\em problem
domain}.
The problem domain is viewed as a part of physical or imaginable
(perceptual) reality, or external world. This is a natural
starting point. As a result the observer is to operate with
a {\em representation}. The represented domain is inhabited
by the (atomic) entities, or {\em individuals}. A safety reason
is to set up individual as a primary concept that is not assumed
to be definable. In fact, the observer operates with the
{\em constructs} that represent the individuals.

%%%%%%% here: {   } is indent %%%%%
\begin{namelist}{{\tt -h} {\it {   }}}
\item {\sf Important}: The possibility does exist to gather the
individuals into a single domain $D$, and this $D$ is
given from the beginning.
\end{namelist}

The advanced studies in a theory of computations prescribe $D$
as a domain of {\em potential} (or: {\em schematic}) individuals.
To the contrast the recent object-oriented studies almost
ignore this fact.
This ignorance does omit namely the feature of potentiality,
or possibility of individual. The individual is possible
with respect to some theory (of individuals).

\begin{namelist}{{\tt -h} {\it {   }}}
\item {\sf Advance}: The individual may be relativized
and gives a family of object-oriented strategies.
\end{namelist}

E.g., `this theory of objects is similar to usual'.
The individuals (theories) enter the domain and leave it
cancelling their own existence. The `flow of events'
in the example may be based on a time flow. Any two
theories are to be compared in spite of their existence in
different
`moments'. The theories are not necessary fixed, thus all
amount of the possible individuals is involved.

\begin{namelist}{{\tt -h} {\it {   }}}
\item {\sf Further advance}: The individuals are separated,
at least, into {\em possible} and {\em virtual}.
\end{namelist}

Only the virtual individuals are completely {\em ideal objects}.
So the regularity of the {\em observer's language} is increased.
In mathematical practice to be a possible individual
means to be {\em described}, but the virtual individual
(objects) does need the {\em axioms}.

\begin{namelist}{{\tt -h} {\it {   }}}
\item {\sf Effect}: The virtual objects increase the structure
regularity of the (initial) domain $D$.
\end{namelist}

As a result, clear distinction between {\em actual, possible}
and {\em virtual} individuals induces the inclusion:
\[ A \subseteq D \subseteq V, \]
where $A$ is a set of actual individuals,
$D$ is a set of possible individuals,
and $V$ is a set of virtual individuals.

\begin{namelist}{{\tt -h} {\it {   }}}
\item {\sf Advance}: The central computational proposal is to
generate actual individuals as the different {\em families} of
$D$,
\[ A_i \subseteq D\ {\rm for\ } i \in I.  \]

\item {\sf Trouble}: The object-oriented approaches propose to
operate
a fuzzy notion of a {\em thing} and {\em property} ignoring
the distinctions between generic and derived concepts.
The language of the observer is likely mixed with the domain $D$.
Thus, the meaning of an individual is violated.
\end{namelist}

\subsection{Other generic notions}

Starting with things and properties the observer builds the
{\em composite things} and establishes for his objects the
attributes (is there any object without attribute ?). Thus,
an observer actually needs a (logical) language, even
overcoming his own initial desire. The obvious approach
is getting started with a choice of logics.

\begin{namelist}{{\tt -h} {\it {   }}}
\item {\sf Trouble}: The logics is not homogeneous.
Its branches, especially for a theory of computations contain the
suitable advances. They do not suit the amorphous idea of a thing
and property.
\end{namelist}

Instead of overcoming this barrier theory of computations enables
the
regular and working logics of {\em the descriptions}.
The descriptions directly illustrate the difficulties
and tend to general operators.

Operating with things and properties gives a specific property -
{\em law}. The law is essentially the constraint superimposing
to the properties of a thing.

Recall that in application the observer assigns attributes
to things (they are not the intrinsic to things
in contrast to properties).

\begin{namelist}{{\tt -h} {\it {   }}}
\item {\sf Important}: Both the
logical formula $\Phi (x)$
and $\lambda-$expression $\lambda x.\Phi(x)$
give the property, but the direct assignment of the property
$\Phi (\cdot)$ to the individual $x$ is given by
the description:
\[ {\cal I}x.\Phi (x),  \]
with a sense `the (unique) $x$ that $\Phi (x)$'
(compare with $\lambda x.\Phi(x)$, `those $x$ that $\Phi(x)$').

\item {\sf Filling in the gap}: The gap between the observer
(and his language) on the one hand and the individuals
on the other hand does exist in object-oriented modelling.
\end{namelist}

An abridgement is given by the {\em evaluation map}:
$$
\parallel \cdot \parallel \cdot\ : \
\left\{
\begin{array}{l}
{\rm descriptions}\\
\lambda-{\rm expressions}
\end{array}
\right\}
\times\ {\rm assignments\ }
\rightarrow\ {\rm individuals}.
$$
(Here: an assignment is temporary viewed as an index
ranging the families.)
The abridged concepts are an {\em attribute} $a$ and
{\em property} $\Phi (\cdot)$ (via the description):
$$
a=\parallel {\cal I}x.\Phi(x)\parallel_i\ {\rm for\ } i\in I
 \eqno({\bf Attr})
$$
An attribute thus defined indicates the set of individuals
with a property $\Phi(\cdot)$. In usual terms the
{\em functional representation of attribute} is established
(attribute is a mapping {\em from} a set of things
and a set of `observation points' {\em into} a set of values).
Note that a `thing' is represented by the `description'.

\begin{namelist}{{\tt -h} {\it {   }}}
\item {\sf Principle adopted}: The attribute is defined
by ({\bf Attr}). The addition of the uniqueness
\[
\begin{array}{lr}
\{ a \} = \{ d \in D \mid\ \| \Phi(\bar{d})\|_i
= true \} \hfill & ({\bf Singleton})
\end{array}
\]
as necessary and sufficient condition
\[
\begin{array}{lr}
\| {\cal I}x.\Phi(x)\|_i = a \Leftrightarrow  \\
\Leftrightarrow \{ a \} =  \{ d \in D \mid\ \| \Phi(\bar{d})\|_i
= true \} \hfill & ({\bf Unique})
\end{array}
\]
enforces the observer to conclude: fixing the family $i\in I$
and evaluating $\|\Phi(\bar{d})\|_i$ relatively to every $d\in
D$,
he verifies the uniqueness of $d$.
\end{namelist}

Here the individual is called as $a$ and is adopted as
an evaluation of the description relatively to $i$.

\subsection{Functional scheme}

A general solution for attributes attracts the set of attribute
functions ({\bf Attr}) that is called as a {\em functional
scheme}.

\begin{namelist}{{\tt -h} {\it {   }}}
\item {\sf Advance}: Equation ({\bf Attr}) is to be revised
as follows:
\[
\begin{array}{ll}
{\cal I}x.\Phi(x) = \bar{h} & {\rm in\ a\ language\ of\ observer}
\vspace{0.5ex}
\\

\|\bar{h}\|=h              & {\rm is\ an\ individual\ concept\ } \\
                           & {\rm in\ a\ domain}
\vspace{0.5ex}
\\

h(i)=a                     & {\rm  is\ an\ individual} \\
                           & {\rm in\ a\ domain}
\end{array}
\]
\item {\sf Further advance}: Previously given scheme has a
universe
of discourse as `concept-individual'. An undevoted observer
if needed may prefer the `individual-state' universe.
\end{namelist}

Thus, if $h$ is an individual, then $a$ is its {\em state}
under the {\em forcing condition} $i$.

\begin{namelist}{{\tt -h} {\it {   }}}
\item {\sf Advantage}: The generalized individuals (or: concepts)
are schematic:
\[ h : I \rightarrow C, \]
where $h$ is a mapping from the `observation points' into
the (subset of) attribute $C$. The latter undoubtedly
is {\em the} set of individuals.
\end{namelist}

There is a clear reason to call $h$ as a {\em concept}.
Thus a concept really represent the functional scheme.

\begin{namelist}{{\tt -h} {\it {   }}}
\item {\sf Effect}: The (individual) functional schemes
are to be gathered into a greater stock:
\[
\begin{array}{lr}
\{h \mid h: I \rightarrow C\} = H_C(I)  \hfill & ({\bf VDom}).
\end{array}
\]
\end{namelist}

Certainly, $H_C(I)$ is and idealized object.

\begin{namelist}{{\tt -h} {\it {   }}}
\item {\sf Important}: The object $H_C(I)$ is
a representation, and what is specific
the feature of a {\em variable domain} is captured.
\end{namelist}

The possibilities and the advantages of a notion of
variable domain are applied mostly to the {\em dynamics}.

\subsection{Dynamics of objects}

The state in an object-oriented approach is viewed as the value
of the functions in the functional scheme at a given point
among the `observation points'. This agrees with the
computational framework.

\begin{namelist}{{\tt -h} {\it {   }}}
\item {\sf Important}: Computationally a set of individuals
is generated by:
\[ H_C(\{ i\} ) \subseteq C\ {\rm for\ } i\in I.  \]
\end{namelist}

This set is a state of a variable domain $H_C(I)$, where
$C$ gives the local universe of possible individuals.
The pointer $i$ marks the family of individuals that
is `observed' from $i$.
The states $s_1, s_2, \ldots$ of a functional scheme
have a representation by the {\em stages} of the variable domain:
\[
\begin{array}{lcl}
H_C(\{ i \}) &=& \{ h(i) \} \subseteq C \\
H_E(\{ i \}) &=& \{ h(i) \} \subseteq E \\
\ldots & & \ldots
\end{array}
\]

Transformations $g : s_1 \mapsto s_2$ are the counterparts
of the {\em events} (they are triples):
\[  <s_1,s_2;g>.  \]

\begin{namelist}{{\tt -h} {\it {   }}}
\item {\sf Generalization}: The notion of a variable domain gives
the natural observation of the dynamics in an object-oriented
approach. Even more, it gives a suitable metatheoretic framework.
\end{namelist}

To cover the possible effects the {\em natural transformations}
$H_g : H_C \rightarrow H_E$ are added. The element-wise
consideration
gives:
\[
\begin{array}{lclcl}
H_g(I) &:& h\in H_C(I) &\mapsto & g\circ h \in H_E(I), \\
H_g(\{ i \}) &:& \{ h(i) \} \subseteq C
                     &\mapsto & (g\circ h)(i) \subseteq E.
\end{array}
\]

\begin{namelist}{{\tt -h} {\it {   }}}
\item {\sf Important}: The set of transformations
gives the {\em laws of things} in object-oriented reasoning.
\end{namelist}

The immediate result gives a clear understanding of interaction
of things (via state variable common to interacting things).
Thus, the set of natural transformations is a representation
of the laws of $\ldots$ . And here is a short diagram of what of:
\[
\begin{array}{c}
\{ h(i) \} \subseteq C \\
x_1 \in \{ h(i) \} ; x_2 \in \{ h(i)\} \\
\ldots\ \Phi(x_1) \& \Psi(x_2) \& x_1=x_2(=z)\ \ldots\ ,
\end{array}
\]
where $z$ is a common variable (joint state variable).

\subsection{Dynamics via evolvent}

The more dynamics may be added to an object. The task under
solution is a {\em behavior} of a thing (= state evolution
`in a time'). Note that the state will change due to both
the external and internal events.

\begin{namelist}{{\tt -h} {\it {   }}}
\item {\sf Important}: The evolvent of stages is needed:
\[ f : B \to I, \]
where stages are evolved {\em from} $I$ {\em to} $B$
(note the reversed order, so $B$ is later than $I$).
\end{namelist}

Computationally are given:  $H_g : H_C \to H_E$ for $g: C\to E$
($C, E$ are the attributes) and $f: B\to I$ for stages $I, B$.
The combined transformation is generated both by $f$ and $g$:
\[
\begin{array}{lcl}
h\in H_C(I)            &\mapsto & g\circ h\circ f \in H_E(B), \\
\{ h(i) \} \subseteq C &\mapsto & ((g\circ h)\circ f)(b)
\subseteq E.
\end{array}
\]
for $b\in B$.

In particular, a stable state is generated by:
\[
\begin{array}{lcl}
f &=& 1_I : I \to I, \\
g &=& 1_C : C \to C.
\end{array}
\]

\subsection{Object characteristics}

The commonly used in object studies are
{\em encapsulation, composition, classification}, and
{\em communication/transaction}.

\begin{namelist}{{\tt -h} {\it {   }}}
\item {\sf Encapsulation}: An object contains: (1)~{\em state},
(2)~capability of {\em transitions} (state changes; actions;
services), and (3)~{\em interface}.
\end{namelist}

Computationally, an object has: (1)~attributes $C, E, \ldots$;
(2)~transformations $g:C\to E,\ \ldots$; and (3)~composable
transformations (possibly, they are closed under composition).
In particular, objects with exclusively {\em interface
attributes}
are viewed as the {\em static objects}. This can be modelled
by $g=1_C:C\to C,\ f=1_I:I\to I$ etc.

\begin{namelist}{{\tt -h} {\it {   }}}
\item {\sf Composition}: As usually, the composite object
is assumed to be combined from the other objects.
\end{namelist}

This means the following: (1) logics (of the properties) is
attached, (2) composition (possibly, in a category) is added etc.
All of this is in a full harmony with the theory of computations.

\begin{namelist}{{\tt -h} {\it {   }}}
\item {\sf Classification}: Traditionally, the objects with
the same set of properties (attributes, actions)
are gathered into a class.
\end{namelist}

The computational generalization attracts the concept
of a variable domain $H_C(I)=\{ h \mid h:I\to C\}$
that is defined over the {\em schematic objects}.

\begin{namelist}{{\tt -h} {\it {   }}}
\item {\sf Communication/interaction}: Ordinarily communication mainly
implies the changes of the object attributes
(change is the same as a {\em request}). A request may cause
a state transition (change of the non-interface attributes;
change the state of the receiver/sender via interface
attributes).
\end{namelist}

\section{Construction of object}

A point of importance to determine an object is the notion of {\em
type}. The known results either illustrate the analogy between
typed and type-free models, or establish their real connection. In
particular, untyped models contain object-as-types via embedding.
The computation, e.g., in type-free $\lambda$-calculus has a goal
to derive an object with the pre-defined properties ({\em dynamic
typing}). To the contrary, the same computation in a typed
$\lambda$-calculus has to obtain the derived type by the rules
({\em static typing}).

To conform types with dynamics they are to be fitted
the dynamical considerations. The initial set of `hard'
types is usually predefined. To the contrary the `soft'
types are derived from the generic to give rise to a more
flexible ground.

Untyped models naturally combine type and its implementation
(embedded objects). Sometimes the researcher may prefer to
separate them.
As a working hypothesis
the thesis `to represent means to classify properly '
meets the opposition from the alternative approach.
This second way tends to the `slight' variations
of the initially formed objects.

\subsection{Embedding objects into functor category}

Give a {\em construction} to accumulate the intuitive
reasons above.
Let to consider more than one category. At first, given
category $\cal C$ is a {\em set} and is assumed as c.c.c.
Let $\cal S$ be the category of all sets and arbitrary
functions, a c.c.c. Construction of the functor category
(it is a c.c.c.) ${\cal S}^{{\cal C}^{op}}$ give all the
(contravariant) functors from $\cal C$ into $\cal S$.
The known result is that the functor category is a model
for higher order logic.

\subsubsection{Object-as-functor}

Let a mapping $F:{\cal C}\to {\cal S}$ be the association
to arbitrary domain $I$ of $\cal C$ a set $F(I)$ of $\cal S$
and to every map $f:B\to I$ of $\cal C$ a function
$F(f):F(I)\to F(B)$ so that:
$$
F(1_I) = 1_{F(I)}, {\rm and} \\
F(f\circ g) = F(g)\circ F(f),
$$
provided $f:B\to I$ and $g:C\to B$ in $\cal C$.

So defined functor $F$ determines the family
of objects parameterized by $I$.

\subsubsection{Object-as-domain}

To construe an object that models the meaning of
the variable domain an example of functor category is used.

For every $T$ of $\cal C$ let
$$
H_T(I)= \{h|h:I\to T\}
$$
and if $f:B\to I$ in $\cal C$, let $H_T(f)$ be the map
taking $h\in H_T(I)$ into $h\circ f \in H_T(B)$.
It is easy to verify $H_T$ is a contravariant functor.

{\em Transactions}. Let $g:T\to S$ in $\cal C$. There is a natural
transformation $H_g:H_T\to H_S$. Every $h\in H_T(I)$  can be
mapped to $g\circ h\in H_S(I)$. So defined mapping $g$ determines
a rectified idea of transaction.

{\em Clones}. The composite map for $f:B\to I$ takes
$h\in H_T(I)$  {\em into} $g\circ h\circ f \in H_S(B)$.
Thus, the individuals from $H_T(I)$ are $f$-cloned
into $H_S(B)$.

It is easy to verify $H:{\cal C}\to {\cal S}^{{\cal C}^{op}}$
is a covariant functor, and $\cal C$ may be assumed to be c.c.c.

\subsubsection{Functorial properties}

Let functor $H_T$ in ${\cal S}^{{\cal C}^{op}}$
be treated as a {\em variable domain}: (1)~for every
$I\in {\cal C}$ an associated domain $H_T(I)$ is the set;
(2)~the maps $f:B\to I$ in $\cal C$ give transitions
from stage $I$ to stage $B$.

Every transition clones elements in $H_T(I)$ into
elements in $H_T(B)$ along the map $f$.

The verification of functorial properties of $H_T$
is straightforward. The properties of the {\em restriction}
come down to the following:
\[
\begin{array}{lcl}
h\rceil 1_I &=& (H_T(1_I))(h) \\
            &=& h\circ 1_I  \\
            &=& h, \\
(h\rceil f)\rceil g &=&h\rceil (f\circ g),
\end{array}
\]
where $h\rceil f = (H_T(f))(h) = h\circ f$ is an
abbreviation.

\section{Fragment of a theory of types}

Many possible theories of types are known, and the need is
of getting down to some details of object-as-functor
for types.

The domains $A$ of $\cal C$ are associated to the {\em type symbols},
and they are basic types. The derived types are generated
by constructions: {\bf 1} (empty product),
$T\times S$ (cartesian product), $T\to S$ (functional space),
$[T]$ (power type).

In the functor category an arbitrary type $T$ is indicated
as $H_T$, and an evaluation mapping $\|\cdot\|$ needs an
additional parameter, so that $\|\cdot\|\cdot$. And this is
an important stage to treat the functor category as
an interpretation for a higher order theory.

\subsection{Dynamics: further understanding via logic}

The construction of a logical framework reflects the
adopted object solutions.

\subsubsection{Logical language}

A language contains a supply of variables for every type.
{\em Atomic formulae} are the equations:
\begin{itemize}
\item[] $x = y$, where $x,\ y$ are of the same type;
\item[] $y = gx$, where $g$ is a constant $g:T\to S$
of $\cal C$, $x$ and $y$ have the types $T$ and $S$
respectively;
\item[] $z = [x, y]$, where $x,\ y$ of types $T,\ S$
respectively, $z$ of type $T\times S$;
\item[] $z = x(y)$, where $x$ has type $T\to S$,
$y$ type $T$, $z$ type $S$;
\item[] $y\in x$, where $y$ is of type $T$,
$x$ of type $[T]$.
\end{itemize}

{\em Formulae} $\Phi$ are generated as usually by the
{\em connectives} and {\em quantifiers}.

\subsubsection{Interpretation}

Assume the following: $A$ is a domain of $\cal C$, $\Phi$
is a formula, $\|\cdot\|$ is an evaluation of the non-bound
variables of $\Phi$.

An evaluation of the {\em variable} makes $\|\cdot\|$ relative to
the domains of $\cal C$ (e.g., to $A$) and needs the explanation.

Visible objects are percepted by the observer via his machinery in
spite of the doctrine of the predefined objects.

The events evolve from $A$ to $B$. The inhabitants of the world
$A$ evolve, so they inhabit the world $B$. The world $B$ contains
the clones of $A$-inhabitants, and also some other inhabitants, if
any.
$$
\begin{CD}
@.                 A          @<f<<           B     \\[3ex]
  \|\bar{x}\|A\in @. H_T(A)     @>H(f)>>   (H_T)_f @>\subseteq>> H_T(B) @. \ni  \|\bar{y}\|B
\end{CD}
$$

\begin{equation}
 \|\bar{x}\|A = \|\bar{y}\| B  = \|\bar{x}\|_fB  \label{eq:var}
\end{equation}

The evaluation of the atomic formulae is getting down
to the case study (are given for atomic case).

{\em Variables}.

$\|x=y\|A \iff \|x\|A = \|y\|A$ \hfill ({\bf Var})%

{\em Constant function}.

$\|y=gx\|A \iff \|y\|A = g\circ\|x\|A$ \hfill ({\bf CFun})%

{\em Ordered pair}.

$\|z=\lbrack x,y\rbrack\|A \iff \|z\|A = \lbrack \|x\|A,\|y\|A\rbrack$
\hfill ({\bf DPair})%

{\em Application (variable function)}.

$\|z=x(y)\|A \iff \|z\|A = \|x\|_{1_A}A(\|y\|A)$ \hfill ({\bf $\varepsilon$})%

{\em Powerset}.

$\|y\in x\|A \iff \|y\|A \in \|x\|_{1_A}A$ \hfill ({\bf PSet(A)})

$\|y\in x\|B \iff \|y\|B \in \|x\|_{1_B}B$ \hfill ({\bf PSet(B)})

$\|y\in x\|_f \iff \|y\|B \in \|x\|_fB$ \hfill ( {\bf PSet${}_f$})

\subsubsection{Construction of concept}

The notion of a `concept' depends on a set of conditions
and was studied under the various assumptions.
The following matches an intuition for a `variable domain'.

A notational remark. In the below $\|\cdot\|_{(t/y)}$ means
the fixed evaluation where $t$ matches $y$ of the same type.
The evaluation $\|\cdot\|_{(t/y)}\rceil f = \|\cdot\|_f$
matches $\| y \|_f$ with every relevant variable $y$.
Any case the restriction $\rceil$ is superimposed to the
functor $H_T$ with $T$ is the type of $y$.

Let concepts $C(A),\ C(B)$, and $C_f$ be the different
restrictions of the $H_T$:

$C(A)= \{t\in H_T(A)\mid\|\Phi(y)\|_{1_A(t/y)}A\}$ \hfill ({\bf Conc(A)})

$C(B)= \{t\in H_T(B)\mid\|\Phi(y)\|_{1_B(t/y)}B\}$ \hfill ({\bf Conc(B)})

$C_f= \{t\in H_T(B)\mid%
% \lbrack t/y_f \rbrack
\|\Phi(y)\|_{f(t/y)}B\}$ \hfill ({\bf Conc${}_f$})

Their relationships correspond to the diagram:

\[
\begin{CD}
           A          @<f<<           B        \\[3ex]
      C(A)           @>C(f)>>      C_f     @>\subseteq>> C(B)
\end{CD}
\]
(here: $C_{1_A}=C(A)$; $C_f \subseteq C(B)$; $C=H_T$ )

\subsection{Case study for variable domains}

The `transaction-clone' notion having been applied to the functor
category $H:{\cal C}\to {\cal S}^{{\cal C}^{op}}$ has a benefit of
explicate arrow-thinking. In the following family of diagrams the
mapping $f:B\to A$ {\em clones} the individual from $A$ into $B$.
Besides that, the mapping $g:C\to D$ represents the {\em
transition} (an explanatory
system is of free choice): \\

{\em general diagram}:

\[
\begin{CD}
\hspace{5em}  @.      A          @<f<<           B \\[3ex]
T             @.      H_T(A)     @>H_T(f)>>   (H_T)_f @>\subseteq>>  H_T(B) \\[1ex]
@VgVV                 @V{H_g(A)}VV                 @.          @VV{H_g(B)}V \\
{\cal T}      @.   H_{\cal T}(A)     @>{H_{\cal T}(f)}>> (H_{\cal T})_f @>\subseteq>>  H_{\cal T}(B)  \\
\end{CD}
\]

singular:
$$H_C(A)=\{h\mid h : A\to [C]\}$$
\[
\begin{CD}
H_C(A)
\end{CD}
\]

$f$-cloned:
$$H_C(f): H_C(A)\ni h \mapsto h\circ f \in H_C(B)$$
\[
\begin{CD}
            A          @<f<<           B   \\[3ex]
           H_C(A)     @>H_C(f)>>      H_C(B)
\end{CD}
\]

non-cloned, $g$-transacted:
$$H_g: H_C(A)\ni h \mapsto g\circ h \in H_D(A)$$

\[
\begin{CD}
\hspace{5em} @.        A       \\[3ex]
C            @.           H_C(A)    \\[1ex]
@VgVV                 @V{H_g(A)}VV  \\[1ex]
D            @.           H_D(A)
\end{CD}
\]

$1_A$-cloned, $g$-transacted:
$$H_D(1_A) \circ H_g: H_C(A)\ni h \mapsto g\circ h\circ 1_A \in H_D(A)$$
\[
\begin{CD}
\hspace{5em} @.            A          @<1_A<<           A   \\[3ex]
C            @.           H_C(A)     @>H_C(1_A)>>      H_C(A)  \\[1ex]
@VgVV                 @V{H_g(A)}VV                @VV{H_g(A)}V \\
D            @.           H_D(A)     @>{H_D(1_A)}>>   H_D(A)
\end{CD}
\]

$f$-cloned, $g$-transacted:
$$H_D(f) \circ H_g: H_C(A)\ni h \mapsto g\circ h\circ f \in H_D(B)$$

\[
\begin{CD}
\hspace{5em} @.            A          @<f<<           B   \\[3ex]
C            @.           H_C(A)     @>H_C(f)>>      H_C(B)  \\[1ex]
@VgVV                 @V{H_g(A)}VV                @VV{H_g(B)}V \\
D            @.           H_D(A)     @>{H_D(f)}>>   H_D(B)
\end{CD}
\]

$f$-cloned, non-transacted:
\begin{eqnarray*}
H_C(f) & : & H_C(A)\ni h \mapsto  h\circ f \in H_C(B)   \\
H_D(f) & : & H_D(A)\ni h \mapsto  h\circ f \in H_D(B)
\end{eqnarray*}

\[
\begin{CD}
             @.            A          @<f<<           B   \\[3ex]
C\hspace{5em}@.      H_C(A)     @>H_C(f)>>      H_C(B)    \\[3ex]
D\hspace{5em}@.   H_D(A_D(A)     @>{H_D(f)}>>   H_D(B)
\end{CD}
\]

The functorial properties of $H_T$ come down
to the case study given above.

\subsection{Evaluation mapping}

The functor category in use may enrich the intuition
concerning an evaluation mapping. In particular,
the diagram given below reflects \\

$f$-cloned evaluation mapping:

\[
\begin{CD}
        A          @<f<<           B   \\[3ex]
\|\Phi\|A    @>\|\Phi\|f>>  \|\Phi\|_f      @>\subseteq>> \|\Phi\|B \\[3ex]
\{y\}         @>H_T(f)>>       \{y\circ f\}     @.              t  \\[1ex]
@V{\in}VV                     @V{\in}VV                    @V{\in}VV \\
H_T(A)        @>H_T(f)>>          (H_T)_f      @>\subseteq>>   H_T(B)
\end{CD}
\]

Similarly, $g$-transacted, $f$-cloned evaluation mappings shown in
Fig.~\ref{Diag:1}.

\begin{figure*}
\[
\begin{CD}
\hspace{5em}  @.      A          @<f<<           B   \\[3ex]
\hspace{5em}  @.  \|\Phi\|A    @>\|\Phi\|f>>  \|\Phi\|_f      @>\subseteq>> \|\Phi\|B \\[3ex]
\hspace{5em}  @.  \{y\}         @>H_C(f)>>       \{y\circ f\}     @.              t  \\[1ex]
\hspace{5em}  @.  @V{\in}VV                         @V{\in}VV                 @V{\in}VV \\
C  @.  H_C(A)        @>H_C(f)>>          (H_C)_f      @>\subseteq>>   H_C(B) \\[1ex]
@VgVV                 @V{H_g(A)}VV        @.        @VV{H_g(B)}V \\
D  @. H_D(A)        @>H_D(f)>>          (H_D)_f      @>\subseteq>>   H_D(B) \\[1ex]
\hspace{5em}  @.  @A{\in}AA                         @A{\in}AA                 @A{\in}AA \\[1ex]
\hspace{5em}  @.  \{g\circ y\}   @>H_D(f)>>  \{(g\circ y)\circ f\}    @. v \\[1ex]
\hspace{5em}  @.    \|               @.           \|         @.      \|  \\
\hspace{5em}  @.  \{u\}   @.     \{u\circ f\}     @.    v  \\[3ex]
\hspace{5em}  @.  \|\Psi\|A    @>\|\Psi\|f>>  \|\Psi\|_f      @>\subseteq>> \|\Psi\|B
\end{CD}
\]
\caption{$g$-transacted, $f$-cloned evaluation mapping}\label{Diag:1}
\end{figure*}
(N.B. Possibly, $\Psi$ may be equal to $\Phi$; $y=u$, and $t=v$.)
The interpretation of previous diagram depends on the
evailable engineering machinery.

An advance in the representation may be achieved with
the {\em concepts} $C_1,\ C_2$ corresponding to
$\Phi,\ \Psi$ respectively.

The previous diagram is comprehenced to:

\[
\begin{CD}
\hspace{5em}  @.      A          @<f<<           B   \\[3ex]
\hspace{5em}  @.  \{y\}         @>{C_1}(f)>>       \{y\circ f\}     @.              t  \\[1ex]
\hspace{5em}  @.  @V{\in}VV                         @V{\in}VV                 @V{\in}VV \\
C  @.  C_1(A)     @>C_1(f)>>     {C_1}_f      @>\subseteq>>   C_1(B) \\[1ex]
@VgVV                 @V{H_g(A)}VV        @.        @VV{H_g(B)}V \\
D  @. C_2(A)      @>C_2(f)>>     {C_2}_f      @>\subseteq>>   C_2(B) \\[1ex]
\hspace{5em}  @.  @A{\in}AA                         @A{\in}AA                 @A{\in}AA \\[1ex]
\hspace{5em}  @.  \{u\}   @>{C_2}(f)>>     \{u\circ f\}     @.              v  \\[3ex]
\end{CD}
\]

The only `transaction-clone' dependencies are visible, so an
explicit object is extracted.

Note in addition, that the concept-image of
$g$-transacted, $f$-cloned evaluation mapping:

\[
\begin{CD}
\hspace{5em}  @.      A          @<f<<           B   \\[3ex]
\hspace{5em}  @.     C_1(A)      @>C_1(f)>>   {C_1}_f     @.         C_1(B)  \\[1ex]
\hspace{5em}  @.  @V{\subseteq}VV     @V{\subseteq}VV      @V{\subseteq}VV \\
C  @.  H_C(A)        @>H_C(f)>>          (H_C)_f      @>\subseteq>>   H_C(B) \\[1ex]
@VgVV                 @V{H_g(A)}VV        @.        @VV{H_g(B)}V \\
D  @. H_D(A)        @>H_D(f)>>          (H_D)_f      @>\subseteq>>   H_D(B) \\[1ex]
\hspace{5em}  @.  @A{\subseteq}AA     @A{\subseteq}AA     @A{\subseteq}AA \\[1ex]
\hspace{5em}  @.     C_2(A)      @>C_2(f)>>   {C_2}_f     @.         C_2(B)
\end{CD}
\]

is in a harmony with the ``logical'' diagram in Fig.~\ref{Diag:1}.

\section{Extracting a computational background}

In applications a theory of functions is based on some
additional objects.

{\em Applicator}
$$
\varepsilon_{BC}:(B\to C)\times B\to C
$$
which applies function $f$ to its argument $x$:
$\varepsilon_{BC}: [f,x] \mapsto f(x)$.

{\em Currying}
$$
\Lambda_{ABC}:(A\times B\to C)\to (A\to (B\to C))
$$
which shifts variables.

More exactly, if $h:A\times B\to C$, then
$\Lambda_{ABC}h:A\to (B\to C)$.

For $k:A\to (B\to C)$ and $h:A\times B\to C$
mapping $\Lambda$ gives a correspondence. Equationally,
it means
\[
\begin{array}{lcl}
\varepsilon \circ <(\Lambda h)\circ Fst, Snd> &=& h, {\rm and} \\
\Lambda(\varepsilon \circ <k \circ Fst, Snd>) &=& k
\end{array}
\]
for the first projection $Fst$ and second projection $Snd$:
$$
Fst: A\times B \to A, \ \ \
Snd: A\times B \to B.
$$
Note, that the equation ($\varepsilon$) may be rewritten:
$$
\|z\| = \varepsilon \circ <\|x\|_{1_A},\|y\|>
$$

Next step will be done to determine the meaning of
an expression.

\subsection{Meaning of expression}

The goal is to determine the meaning of an expression
$F(x)$, or $Fx$ where $F$ is the description of a function
and $x$ is a formal parameter. Thus, $x$ is bound,
or substitutional variable.

A treatment may be simplified with the $\lambda$-notations.
The expression above is to be denoted as $\lambda x.yx$ where
the description of a function $F$ is associated to a variable $y$.

The meaning of a function depends on the meanings of its sub-parts
$y,\ x,\ yx$. Those components, in turn, depend on the value of $y$.

\subsubsection{Building an access}

The values of the variables are available via {\em access
functions} from an {\em environment}. The representation of an
environment is given by the domains $D_y,\ D_x,\ \dots$ which are
the ranges of possible values of $y,\ x,\ \dots$. The domains
$D_y,\ D_x$ give the explicit part of an environment $Env$, and
its implicit rest (not be detailed for the current consideration)
is denoted by $E$:
$$
Env = (E\times D_y)\times D_x
$$

\subsubsection{Case study}

{\em Atomic parts}. An object $\lambda x.yx$  contains atoms
$y,\ x$, and non-atomic part $yx$:
$$
\|y\|:Env\to D_y,\ \ \ \ \ \|x\|:Env\to D_x.
$$

{\em Non-atomic parts}. A non-atomic part $yx$ is evaluated
as follows:
\begin{itemize}
\item[] the pair $<\|y\|,\|x\|>$ is composed, and
$$
<\|y\|,\|x\|> :Env\to D_y\times D_x;
$$
\item[] the metaoperator $\varepsilon$ is applied to the pair:
$$
\varepsilon \circ <\|y\|,\|x\|>.
$$
\end{itemize}

To exemplify let $D_y = (D_x \to D'_y)$; thus, $\varepsilon:
{(D'_y)}^{D_x} \times D_x \to D'_y$ is determined by
$\varepsilon[u,v] = u(v) = uv$, and $D'_y$ is the range for
$\|yx\|, i.e.$
$$
\|yx\| = \varepsilon \circ <\|y\|,\|x\|> :Env\to D'_y
$$

\subsubsection{Substitution}

The expression $\lambda x.yx$ contains $y$ (free variable)
and does not contain $x$ (bound, or substitutional variable;
$x$ may be renamed, if needed). To take into account this
reason the modified environment $Env\times D_x$ is temporary
generated to support the substitution $Subst_x$:
$$
Subst_x : Env\times D_x \to Env,
$$
where for $i\in Env,\ h'\in D_x$ the result is
$$
Subst_x[i,h'] = i_{(h'/x)}.
$$
It means that substitution $Subst_x$ for every ordered pair
$[i,h']$ gives a correspondent environment $i_{(h'/x)}$
which differs from i exclusively in a point $x$
($x$ is substituted by $h'$).

An {\em access function} for $Subst_x$ is generated
by the equation:
$$
Subst_x = <Fst\circ Fst, Snd>
$$

\subsubsection{Composition}

An observation is as follows: the function $\|yx\|$
and $Subst_x$ are composed:
$$
\|yx\| \circ Subst_x : Env \times D_x \to D'_y
$$

The meaning of $\lambda x.yx$ depends on $Env$ for $y$ ($y$ has a
free occurrence in $\lambda x.yx$, and $x$ is bound). Thus,
$\|\lambda x.yx\|$ is a function that associate to $y$ the
function associating $yx$ to $x$. A type consideration gives:
$$
\|\lambda x.yx\| : Env \to {(D'_y)}^{D_x}
$$

To the contrast $\|yx\|$ is a function {\em from}
$(E\times D_y)$ and $D_x$:
$$
\|yx\| : (E\times D_y)\times D_x \to D'_y
$$

Some difficulties exist to establish the correspondence
between meanings $\|\lambda x.yx\|,\ \|yx\|,\ Subst_x$.

\subsection{Correspondence of the meanings}

Let $\|yx\| \circ Subst_x = g$, and
$$
g([i,h']) \in D'_y
$$
for $g: Env \times D_x \to D'_y$.

For $i\in Env$ and every $h'\in D_x$ the function
$g$ is determined by
$g_i(h') = g([i,h'])$. Now the function $\hat{g}$ is defined
by the equation $\hat{g}(i) = g_i$ for $h'\in D_x$. For arbitrary
pair $[i,h']\in Env\times D_x$ the equation
$$
\varepsilon[\hat{g}(i),h'] = g_i(h') = g([i,h'])
$$
is valid.

Note, that an operation $\hat{\cdot}$ generates the additional
metaoperator $\Lambda$ of {\em currying}:
$$
(\Lambda(g)(i))(h') = g([i,h'])
$$

Hence, a curried version of $g = \|yx\| \circ Subst_x$ is
exactly $\|\lambda x.yx\|$, and finally the needed equation
is obtained:
$$
\|\lambda x.yx\| = \Lambda(\|yx\| \circ Subst_x)
$$

Let to summarize the above reasons in Fig.~\ref{pic:A}.
\begin{figure}
\footnotesize
\unitlength=1.00mm \special{em:linewidth 0.4pt}
\linethickness{0.4pt}
\begin{picture}(65.00,29.00)
\put(19.00,27.00){\makebox(0,0)[cc]{$Env\times D_x$}}
\put(19.00,25.00){\vector(0,-1){17.00}}
\put(19.00,4.00){\makebox(0,0)[cc]{${(D'_y)}^{D_x}\times D_x$}}
\put(8.00,17.00){\makebox(0,0)[cc]{$\hat{g}\times id_{D_x}$}}
\put(42.00,29.00){\makebox(0,0)[cc]{$g$}}
\put(23.00,6.00){\vector(2,1){38.00}}
\put(39.00,17.00){\makebox(0,0)[cc]{$\varepsilon$}}
\put(65.00,27.00){\makebox(0,0)[cc]{$D'_y$}}
\put(26.00,27.00){\vector(1,0){35.00}}
\end{picture}
\caption{Commutative diagram for $\|yx\|\circ
Subst_x=g$}\label{pic:A}
\end{figure}
In this figure the following notations are used:
\[
\begin{array}{l}
g: Env \times D_x \to D'_y,\ i\in Env,\ h'\in D_x \\
g_i: D_x \to D'_y,\ g_i(h') = g([i,h']),\ \hat{g}(i) = g_i \\
\lbrack i,h'\rbrack \in Env \times D_x \\
\varepsilon([\hat{g}(i),h']) = g_i(h') = g([i,h'])
\end{array}
\]

At last, an access function for $\|\lambda x.yx\|$ is generated
in accordance with the equation:
$$
\|\lambda x.yx\| =
  \Lambda ((\varepsilon \circ <Snd\circ Fst, Snd>)
            \circ <Fst\circ Fst, Snd>)
$$

It is easy to verify an optimized version of the access function:
$$
\|\lambda x.yx\| =
  \Lambda (\varepsilon \circ <Snd\circ Fst\circ Fst, Snd>)
$$
from the properties of pairs $<\cdot , \cdot>$ and composition.

\subsection{Examples}

Some examples of computation are briefly given below.

{\em Constant $c$}.

\begin{tabbing}
999 \= $= \|0!\|[i,c'] = $ \= $i\in Env,\ c'\in\{ c\}$ for  \kill
1 \> $\|c\|i =$       \> $i\in Env,\ c'\in\{ c\}$ for singleton $\{c\}$    \\
2 \> $= \|0!\|[i,c']$ \> $\|0!\|$ -- a.f. to $\{c\}$ in $Env$ \\
3 \> $= Snd[i,c']$                                          \\
4 \> $= c'(= c)$
\end{tabbing}

{\em Variable $x$}. The evaluation of a variable gives
   one of the possible atomic cases. The abbreviations
   $F$ for $Fst$ and $S$ for $Snd$ are used.

\begin{tabbing}
999 \= $= (\|x\|\circ Subst_x)[i,h']$ www \=  \kill
1   \> $\|x\|i =$               \> $i\in Env$ \\
\hspace{5em} Generation of a.f. : \\
    \> \ \ $= \|0!\|i_{(h'/x)}$     \> $h'\in D_x$; \\
    \>                          \> $\|0!\|$ --  a.f. to $D_x$ in $Env$ \\
    \> \ \ $= \|0!\|[i,h']$         \> $Env = E \times D_x$ \\
    \> \ \ $= S[i,h']$  \\
    \> \ \ $= h'$       \\
2   \> $= (\|x\|\circ Subst_x)[i,h']$  \> $Subst_x:Env\times D_x\to Env$ \\
    \>                                 \> $Subst_x=<F\circ F,S>$  \\
3   \> $= (S\circ <F\circ F,S>)[i,h']$ \> Replace by a.f. \\
4   \> $= S(<F\circ F,S>[i,h'])$         \> Substitution \\
5   \> $= S[F(i),h']$                    \> a.f.         \\
6   \> $= h'$                            \> $h'\in D_x$
\end{tabbing}

{\em Identity transformation}. The evaluation of an identity
  transformation gives a clear separation of access functions
  (a.f.) and substitution.

\begin{tabbing}
999 \= $= (\|x\|\circ Subst_x)[i,h']$ www \=  \kill
1  \> $\|(\lambda x.x)h\|i$=       \> $i\in Env$ \\
2  \> $= \|\lambda x.x\|ih'$          \> $h'\in D_x$ \\
\hspace{5em} Generation of direct access: \\
   \> \ \ \ $= \Lambda\|0!\|ih'$    \> $\|0!\|$ -- a.f. to $D_x$ in $Env$, \\
   \>                           \> $x$ -- bound variable, \\
   \>                           \>  $Env=E\times D_x$ \\
   \> \ \ \ $=S[i,h']$          \> $[i,h']\in Env\times D_x$ \\
   \> \ \ \ $= h'$                                           \\
3  \> $= \Lambda(\|x\|\circ Subst_x)ih'$ \> Using a.f. \\
4  \> $=\Lambda(S\circ<F\circ F,S>)ih'$  \> $Subst_x:Env\times D_x\to Env$ \\
   \>                                 \> $Subst_x=<F\circ F,S>$  \\
5  \> $= (S\circ <F\circ F,S>)[i,h']$ \> $[i,h']\in Env\times D_x$ \\
6  \> $= S(<F\circ F,S>[i,h'])$       \> Substitution \\
7  \> $= S[F(i),h']$                  \> a.f.  \\
8  \> $h'$
\end{tabbing}

{\em Compound evaluation}.

\begin{tabbing}
999 \= %  $=(\varepsilon\circ<S\circ F,S>\circ<F\circ F,S>)[i,h']$
         mmmmmmmmmmmmmm   \=  \kill
1  \> $\|(\lambda x.fx)h\|i =$       \> $i\in Env$ \\
2  \> $= \|(\lambda x.fx)\|ih'$        \> $h'\in D_x$ \\
\hspace{5em} Generation of access: \\
   \> \ \ $= \Lambda\|f0!\|ih'$    \> $\|0!\|$ -- a.f. to $D_x$ in $Env$ \\
   \>                              \> $Env=(E\times D_f)\times D_x$ \\
   \> \ \ $= \|f0!\|[i,h']$        \> $[i,h']\in Env\times D_x$ \\
   \> \ \ $= (\varepsilon\circ<\|f\|,S>)[i,h']$ \> $S(i)\in D_x$ \\
   \> \ \ $=\varepsilon[\|f\|[i,h'],h']$
                          \> $\|f\|$ -- a.f. to $D_f$ in $Env\times D_x$, \\
   \>                     \> i.e. $\|f\| = S\circ F\circ F$ \\
   \> \ \ $=\varepsilon[(S\circ F)(i),h']$  \> $(S\circ F)(i)\in D_f$ \\
   \> \ \ $= f'h'$  \\
3  \> $=\Lambda(\|fx\| \circ Subst_x)ih'$  \> Replace by a.f.  \\
4  \> $= \Lambda((\varepsilon\circ<S\circ F,S>)\circ<F\circ F,S>)ih'$ \\
   \> (for $Subst_x=<F\circ F,S>,\ \ Subst_x:Env\times D_x\to Env$) \\
5  \> $= (\varepsilon\circ<S\circ F,S>\circ<F\circ F,S>)[i,h']$ \\
   \> (for $[i,h']\in Env\times D_x$) \\
6  \> $= (\varepsilon\circ<S\circ F,S>)(<F\circ F,S>[i,h'])$ \\
   \> (Substitution) \\
7  \> $= (\varepsilon\circ<S\circ F,S>)[F(i),h']$ \\
   \> (for $F(i)\in E\times D_f,\ \ h'\in D_x$) \\
8  \> $= \varepsilon[<S\circ F,S>[F(i),h']]$ \> a.f. \\
9  \> $= \varepsilon[(S\circ F)(i),h']$  \> $\varepsilon$; \\
   \> ($(S\circ F)(i)$ extracts value of $D_f$) \\
10 \> $= \varepsilon[f',h']$ \\
11 \> $= f'h'$
\end{tabbing}

\subsection{Advanced examples}

The additional examples of generalized nature involve
more complicated objects.

{\em Evaluation of formula}.
This kind of object has the following equations:

\begin{tabbing}
$\Lambda\|\Phi\|i(hi)$ \ \ \ \= =  $\Lambda\|\Phi\|(Fst[i,hi])(Snd[i,hi])$ \kill
$\|\Phi\|[i,hi]$       \> $= \Lambda\|\Phi\|(Fst[i,hi])(Snd[i,hi])$ \\
$\Lambda\|\Phi\|i(hi)$ \> $= \Lambda\|\Phi\|(Fst[i,hi])(Snd[i,hi])$ \\
                       \> $= \varepsilon[\Lambda\|\Phi\|(Fst[i,hi]),(Snd[i,hi])]$ \\
                       \> $= (\varepsilon\circ<\Lambda\|\Phi\|\circ Fst,id\circ Snd>)[i,hi]$ \\
                       \> $= \|\Phi\|[i,hi]$  \\
 $\|\Phi\|$          \> $= \varepsilon\circ<\Lambda\|\Phi\|\circ Fst,id\circ Snd>$
\end{tabbing}

An abbreviation
$$
\|\Phi\| = \|\Phi(x)\| \circ Subst_x
$$
is used if there is no ambiguity. Hereafter $T$ is a type of
substitutional variable $x$, and environment $Env$ is renamed by
$I$.

\begin{figure}
%\parbox{4cm}{
\footnotesize
\unitlength=1.00mm \special{em:linewidth 0.4pt}
\linethickness{0.4pt}
\begin{picture}(65.00,32.00)
\put(23.00,30.00){\makebox(0,0)[cc]{$I\times T$}}
\put(65.00,30.00){\makebox(0,0)[cc]{[~~]}}
\put(42.00,32.00){\makebox(0,0)[cc]{$\|\Phi\| [\cdot,\cdot]$}}
\put(23.00,28.00){\vector(0,-1){19.00}}
\put(23.00,6.00){\makebox(0,0)[cc]{$[T]\times T$}}
\put(14.00,19.00){\makebox(0,0)[cc]{$\Lambda\|\Phi\|\cdot\times
id_T$}} \put(42.00,19.00){\makebox(0,0)[cc]{$\varepsilon$}}
\put(26.00,8.00){\vector(2,1){39.00}}
\put(27.00,30.00){\vector(1,0){35.00}}
\end{picture}
%\vspace*{40mm}
\caption{Evaluation in c.c.c.}\label{pic:2}
%}
\end{figure}
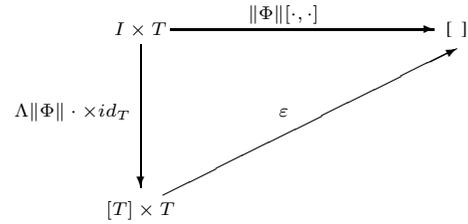

{\em Evaluation in c.c.c.}
The diagram in Fig.~\ref{pic:2} illustrates an idea.

\begin{itemize}
\item  $\Lambda\|\Phi\| : I \to \lbrack T \rbrack$;
\item $\|\Phi\| : I \times T \to \lbrack~~\rbrack$;
\item For $i\in I$ and $hi\in T$ an evaluation
$\varepsilon \lbrack\Lambda\|\Phi\|i, id_T(hi)\rbrack$
generates the truth values from $[~]$.
\end{itemize}

{\em Individuals in c.c.c.}~%
A correspondence of the distinct forms of individuals shows
their similarities.

\fbox{$R\leadsto h_R$.} Given the relation $R\subseteq I\times T$ a function
$h_R : I \to [T]$
is determined by the equality
$h_R(i) =\{h'\mid h'\in T \land iRh'\}$.
In fact, this defines the correspondence $R\leadsto h_R$.~%
\rule{1ex}{1ex}

\fbox{$h\leadsto R_h$.} Given the sets $I,\ T$
the bijection between functions
from $I$ into $[T]$ and the relations from $I$ to $T$
is defined as follows.
The function $h:I\to [T]$ determines the relation $R_h\subseteq I\times T$
by the biconditional $iR_hh'' \iff h''\in h(i)$
for $i\in I$ and $h''\in T$.~%
\rule{1ex}{1ex}

\fbox{${\in}_T$.} The domain
$\in_T = \{ <U,h'>\mid U\subseteq T,\ h'\in T \land h'\in U\}$
is the relation containing all the necessary information
concerning element-subset inclusions.
The following biconditionals are valid:
\[
[i,h']\in R \iff h'\in h_R(i) \iff [h_R(i),h'] \in {\in}_T
\]
Hence, $R$ is a domain and $\in_T$ is a range for mapping
$h_R\times 1_T$ where
$h_R\times 1_T : [i,h']\mapsto [h_R(i),h']$.~%
\rule{1ex}{1ex}

\begin{figure}[t]
%\parbox{4cm}{
\footnotesize
\unitlength=1.00mm \special{em:linewidth 0.4pt}
\linethickness{0.4pt}
\begin{picture}(54.00,32.00)
\put(22.00,6.00){\oval(2.00,2.00)[l]}
\put(18.00,30.00){\makebox(0,0)[cc]{$R$}}
\put(18.00,5.00){\makebox(0,0)[cc]{$\in_T$}}
\put(51.00,5.00){\makebox(0,0)[lc]{$[T]\times T$}}
\put(51.00,30.00){\makebox(0,0)[lc]{$I\times T$}}
\put(53.00,28.00){\vector(0,-1){20.00}}
\put(14.00,17.00){\makebox(0,0)[cc]{$g$}}
\put(54.00,17.00){\makebox(0,0)[lc]{$h_R\times id_T$}}
\put(18.00,28.00){\vector(0,-1){20.00}}
\put(22.00,5.00){\vector(1,0){26.00}}
\put(22.00,31.00){\oval(2.00,2.00)[l]}
\put(22.00,30.00){\vector(1,0){26.00}}
\end{picture}
%\vspace*{40mm}
\caption{Variants of individuals}\label{pic:3}
%}
\end{figure}
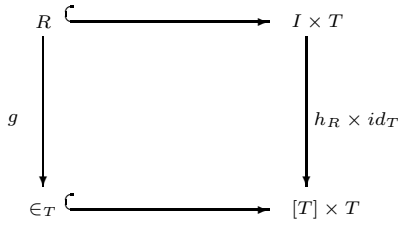

The diagram in Fig.~\ref{pic:3} reflects the ideas given above.

Here: $g$ is an $R$-restricted version of $h_R\times id_T$.
Note that all of this is quite elementary.

\begin{figure}[t]
\footnotesize
\unitlength=1.00mm \special{em:linewidth 0.4pt}
\linethickness{0.4pt}
\begin{picture}(73.00,32.00)
\put(18.00,31.00){\oval(2.00,2.00)[l]}
\put(18.00,6.00){\oval(2.00,2.00)[l]}
\put(14.00,30.00){\makebox(0,0)[cc]{$R$}}
\put(14.00,5.00){\makebox(0,0)[cc]{$\in_T$}}
\put(36.00,5.00){\makebox(0,0)[lc]{$[T]\times T$}}
\put(36.00,30.00){\makebox(0,0)[lc]{$I\times T$}}
\put(38.00,28.00){\vector(0,-1){20.00}}
\put(10.00,17.00){\makebox(0,0)[rc]{$C\cdot\times id_T$}}
\put(14.00,28.00){\vector(0,-1){20.00}}
\put(73.00,30.00){\makebox(0,0)[cc]{[~~]}}
\put(53.00,32.00){\makebox(0,0)[cc]{$\|\Phi\| [\cdot,\cdot]$}}
\put(27.00,17.00){\makebox(0,0)[cc]{$\Lambda\|\Phi\|\cdot\times
id_T$}} \put(50.00,17.00){\makebox(0,0)[cc]{$\varepsilon$}}
\put(18.00,30.00){\vector(1,0){16.00}}
\put(18.00,5.00){\vector(1,0){16.00}}
\put(42.00,8.00){\vector(4,3){26.00}}
\put(45.00,30.00){\vector(1,0){23.00}}
\end{picture}
\caption{Computational properties}\label{pic:4}
\end{figure}
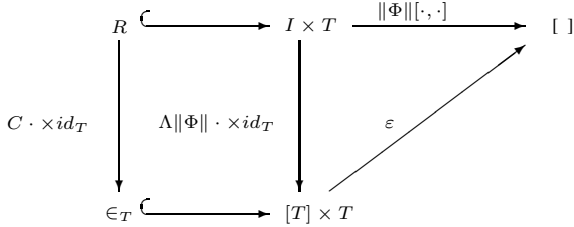

{\em Computational properties of the individuals}. The combined
diagram in Fig.~\ref{pic:4} establishes not so evident
correspondences. What is important that the functor $C\cdot\times
id_T$ includes as a left counterpart the mapping $C\cdot : I\to
[T]$. This mapping is relative to relation $R$ and this relation
is induced by the evaluation of the restriction $\Phi$.

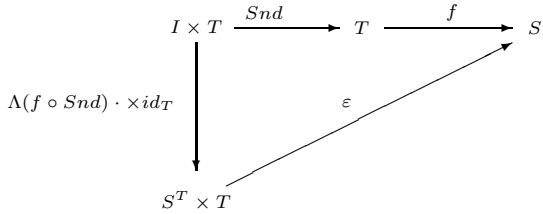
\begin{figure}[t]
\footnotesize
\unitlength=1.00mm \special{em:linewidth 0.4pt}
\linethickness{0.4pt}
\begin{picture}(71.00,27.00)
\put(26.00,25.00){\makebox(0,0)[cc]{$I\times T$}}
\put(48.00,25.00){\makebox(0,0)[cc]{$T$}}
\put(26.00,23.00){\vector(0,-1){17.00}}
\put(26.00,2.00){\makebox(0,0)[cc]{$S^T\times T$}}
\put(12.00,15.00){\makebox(0,0)[cc]{$\Lambda(f\circ
Snd)\cdot\times id_T$}} \put(71.00,25.00){\makebox(0,0)[cc]{$S$}}
\put(35.00,27.00){\makebox(0,0)[cc]{$Snd$}}
\put(60.00,27.00){\makebox(0,0)[cc]{$f$}}
\put(30.00,4.00){\vector(2,1){38.00}}
\put(46.00,15.00){\makebox(0,0)[cc]{$\varepsilon$}}
\put(31.00,25.00){\vector(1,0){14.00}}
\put(51.00,25.00){\vector(1,0){17.00}}
\end{picture}
\caption{Built-in function $f$}\label{pic:5}
\end{figure}

In particular, a built-in function for the given (and evaluated)
argument in a category results in the diagram in Fig.~\ref{pic:5}.

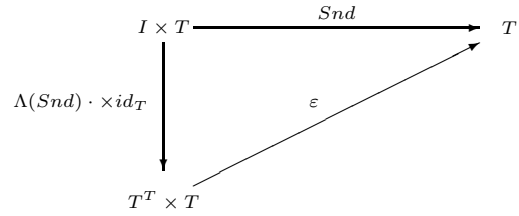
\begin{figure}[t]
\footnotesize
\unitlength=1.00mm \special{em:linewidth 0.4pt}
\linethickness{0.4pt}
\begin{picture}(65.00,29.00)
\put(19.00,27.00){\makebox(0,0)[cc]{$I\times T$}}
\put(19.00,25.00){\vector(0,-1){17.00}}
\put(19.00,4.00){\makebox(0,0)[cc]{$T^T\times T$}}
\put(8.00,17.00){\makebox(0,0)[cc]{$\Lambda(Snd)\cdot\times
id_T$}} \put(42.00,29.00){\makebox(0,0)[cc]{$Snd$}}
\put(23.00,6.00){\vector(2,1){38.00}}
\put(39.00,17.00){\makebox(0,0)[cc]{$\varepsilon$}}
\put(23.00,27.00){\vector(1,0){38.00}}
\put(65.00,27.00){\makebox(0,0)[cc]{$T$}}
\end{picture}
\caption{Free variable}\label{pic:6}
\end{figure}

A free variable is evaluated according to the diagram
in Fig.~\ref{pic:6}.

\begin{figure*}[t]
\footnotesize
\unitlength=1.00mm \special{em:linewidth 0.4pt}
\linethickness{0.4pt}
\begin{picture}(144.00,88.00)
\put(40.00,30.00){\vector(0,-1){20.00}}
\put(47.00,32.00){\vector(1,0){26.00}}
\put(85.00,32.00){\vector(1,0){28.00}}
\put(44.00,8.00){\vector(3,1){69.00}}
\put(118.00,32.00){\makebox(0,0)[cc]{$T$}}
\put(118.00,37.00){\makebox(0,0)[cc]{$5$}}
\put(79.00,32.00){\makebox(0,0)[cc]{$T\times T$}}
\put(79.00,37.00){\makebox(0,0)[cc]{$[2, 3]$}}
\put(38.00,32.00){\makebox(0,0)[cc]{$I\times(T\times T)$}}
\put(38.00,4.00){\makebox(0,0)[cc]{$T^{T\times T}\times(T\times
T)$}} \put(24.00,20.00){\makebox(0,0)[cc]{$\Lambda(+\circ
Snd)\cdot\times id_{T\times T}$}}
\put(72.00,20.00){\makebox(0,0)[cc]{$\varepsilon$}}
\put(60.00,34.00){\makebox(0,0)[cc]{$Snd$}}
\put(99.00,34.00){\makebox(0,0)[cc]{$+$}}
\put(73.00,53.00){\makebox(0,0)[cc]{$T^T\times T$}}
\put(86.00,53.00){\makebox(0,0)[cc]{$T^T\times T$}}
\put(58.00,55.00){\makebox(0,0)[cc]{$\varepsilon$}}
\put(117.00,55.00){\makebox(0,0)[cc]{$\varepsilon$}}
\put(18.00,53.00){\makebox(0,0)[cc]{$T$}}
\put(142.00,53.00){\makebox(0,0)[cc]{$T$}}
\put(144.00,58.00){\makebox(0,0)[cc]{$3$}}
\put(17.00,58.00){\makebox(0,0)[cc]{$2$}}
\put(46.00,67.00){\makebox(0,0)[cc]{$T$}}
\put(114.00,67.00){\makebox(0,0)[cc]{$T$}}
\put(116.00,72.00){\makebox(0,0)[cc]{$3$}}
\put(43.00,72.00){\makebox(0,0)[cc]{$2$}}
\put(77.00,79.00){\vector(0,-1){23.00}}
\put(83.00,79.00){\vector(0,-1){23.00}}
\put(74.00,82.00){\makebox(0,0)[cc]{$I\times T$}}
\put(86.00,82.00){\makebox(0,0)[cc]{$I\times T$}}
\put(74.00,88.00){\makebox(0,0)[cc]{$[i,2]$}}
\put(86.00,88.00){\makebox(0,0)[cc]{$[i,3]$}}
\put(57.00,76.00){\makebox(0,0)[cc]{$Snd$}}
\put(103.00,76.00){\makebox(0,0)[cc]{$Snd$}}
\put(32.00,64.00){\makebox(0,0)[cc]{$can_T$}}
\put(128.00,64.00){\makebox(0,0)[cc]{$can_T$}}
\put(60.00,60.00){\makebox(0,0)[cc]{$\Lambda(can_T\circ
Snd)\cdot\times id_T$}}
\put(99.00,60.00){\makebox(0,0)[cc]{$\Lambda(can_T\circ
Snd)\cdot\times id_T$}} \put(91.00,79.00){\vector(2,-1){19.00}}
\put(117.00,66.00){\vector(2,-1){21.00}}
\put(83.00,34.00){\vector(4,1){56.00}}
\put(96.00,53.00){\vector(1,0){42.00}}
\put(102.00,42.00){\makebox(0,0)[cc]{$Snd$}}
\put(65.00,53.00){\vector(-1,0){42.00}}
\put(70.00,79.00){\vector(-2,-1){20.00}}
\put(44.00,66.00){\vector(-2,-1){19.00}}
\put(75.00,34.00){\vector(-4,1){54.00}}
\put(54.00,42.00){\makebox(0,0)[cc]{$Fst$}}
\put(50.00,88.00){\makebox(0,0)[cc]{entry{\_}point 1}}
\put(110.00,88.00){\makebox(0,0)[cc]{entry{\_}point 2}}
\end{picture}
\caption{An example of computation $+[2,3]$}\label{pic:7}
\end{figure*}
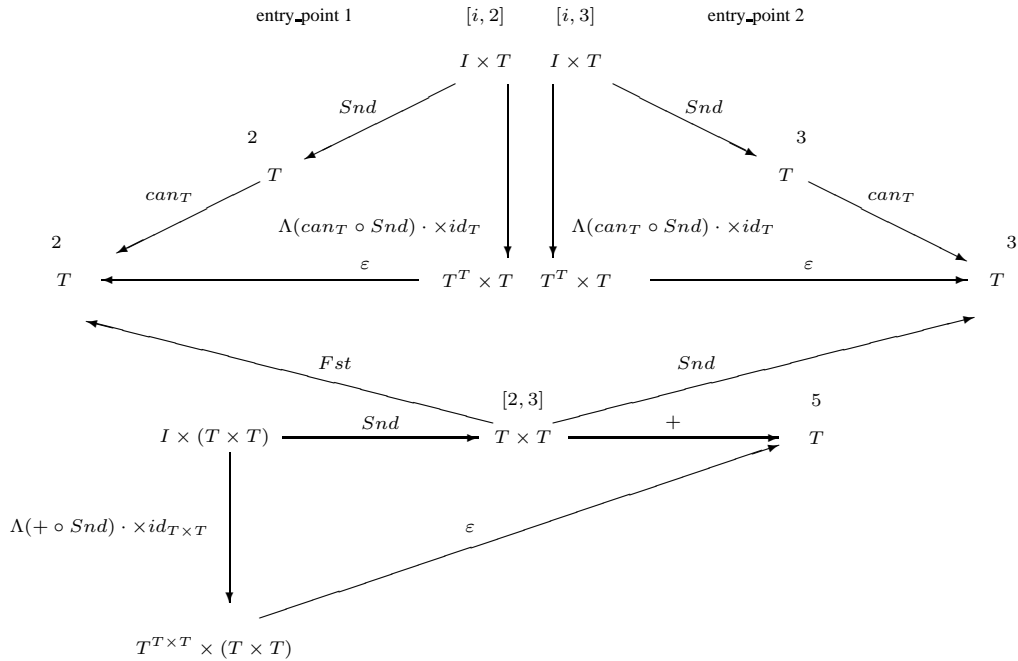

A simplified example of computation (note that both the operands
are to be embedded into the computational environment) like
$$
+[[2/x_1]x_1,[3/x_2]x_2]
$$
is in Fig.~\ref{pic:7}. The entry
points for the computations of the distinct operands are, in
general, independent. Thus, both the left-part and right-part
computations are to be started at the same `moment'. An additional
mappings $can_T$ of canonical embedding of the constants are also
used.

The more exact correspondences are as follow:
\[
\begin{array}{lcl}
R &=& \{ [i,h']\mid \|\Phi\| [i,h']=1\}  \\
\Lambda\|\Phi\|i &\in& [T] \\
\Lambda\|\Phi\| &:&I\to [T] \\
C(\{i\})&=&\{h(i)\mid \Lambda\|\Phi\|i(hi)=1 \} \\
C(I) &=& \{ h\mid\|\Phi(x)\|_{[h/x]} : I\to [T]\}
\end{array}
\]
(here: $x, h:I\to [T]$, so $h(i)\subseteq T$; $x$ is a free variable.)

% \begin{figure}
% %.\special{em:graph W_02.eps}
% \special{psfile picture.ps}
% \vspace*{9cm}
% %\epsffile{c:/tex/tis/work/w_02.eps \the\textwidth}
% \caption{Пример растрового файла.}\label{pic:2}
% \end{figure}

\section{Conclusions}

A common object technique shared by distinct `dimensions' --
logical, categorical, and computational is outlined.

\begin{namelist}{{\tt -h} {\it {   }}}
\item {\sf Important}: The notion of a variable domain
gives a sound ground of the communication analysis
(see, e.g.: \cite{WaWoo:94}, \cite{Jacob:92}).
As may be shown they generate the specific diagrams to
consider the variety of transition effects.
\item {\sf Open discussion}: The questions arise:  \\
1. Is the language of categories adequate to database dynamics
even though the object-oriented approach successively applied?
\end{namelist}
%   \nocite{Bro:94}
%\nocite{Brod:95}
% \nocite{NeRo:95}
%\nocite{EhGoSe:93}

%%%%%%%%%%%%%%%%%%%%% Bibliography %%%%%%%%%%%%%%%%%%%%%%%%%%%%%%%

\addcontentsline{toc}{section}{References}
%\bibliography{bbl_ad95,bibl_00}
\newcommand{\noopsort}[1]{} \newcommand{\printfirst}[2]{#1}
  \newcommand{\singleletter}[1]{#1} \newcommand{\switchargs}[2]{#2#1}

%%%%%%%%%%%%%%%%%%%%%%%%%%%%%%%%%%%%%%%%%%%%

\end{document}